\documentclass[letter]{aa}  

\usepackage{graphicx}
\usepackage{txfonts}
\usepackage{hyperref}
\usepackage[squaren]{SIunits}

\newcommand{\Msun}{\mathrm{M}_{\sun}}				% unit: solar mass

\title{Unexpected correlations in outstanding subpopulations of the gravitational wave transient catalogue data}
\titlerunning{Unexpected correlations in outstanding subpopulations of GWTC3}
\author{Matthias U. Kruckow \and Zhanwen Han}%\inst{1}	%, other Coauthors (Who?),
\authorrunning{Kruckow~et~al.}
\institute{Yunnan Observatories, Chinese Academy of Sciences, Kunming 650216, China\\\email{mkruckow@ynao.ac.cn}}%\and
\date{Received ?? ??, 2021; accepted ?? ??, ????}
\abstract
   {With the third observing run the number of gravitational wave emitting events has increased significantly.}
   {The data of the recorded events is inspected to search for overall properties on the population.}
   {The properties of subpopulations are determined and compared to predictions from simulations.}
   {It appears that the most outstanding systems follow linear relations in the parameter space of the total binary and the chirp mass. Those relations are too tight to have a stochastic origin and are supported by at least five independent events each.}
   {The origin of the correlations is still open to be confirmed, while possible sources, ranging from instrumental artefacts to unknown physics, are discussed and partly excluded. Depending on the relation's source the two events (GW190814 and GW200210\_092254) having a smaller mass component between 2.5 and \unit{2.9}{\Msun} may reveal in a very different light.}

\begin{document} 

\maketitle
%
%-------------------------------------------------------------------

\section{Introduction}\label{sec:Introduction}
The Gravitational-Wave Transient Catalogs (GWTCs) reported in \citet{GWTC1,GWTC2,GWTC3} provide an excellent tool to investigate the population of observed events of coalescing binaries consisting of compact objects. The number of events with determined parameters of the progenitors increased significantly during the third observing run of the ground based gravitational wave detector network. While the first observing run enabled us to observe the mergers of two merging black holes, the second one provided the first double neutron star merger. With the merger of neutron star+black hole mergers all the three observable categories of isolated binary mergers are observed by us.

The formation of those sources is widely studied \citep[for a summary see][ and references therein]{mb21}. The most common origin of two merging compact objects is isolated binary evolution or within higher order multiple stellar systems. The formation with importance of dynamical interactions in dense stellar environments like clusters maybe responsible for a smaller part of the events. A nearly unknown amount of merging events may result from a primordial black hole population which is inaccessible to electromagnetic observations and therefore unconfirmed to exist. For the gravitational wave merging observations it is difficult to clearly differentiate the formation path.

%Tbc.

In Sect.~\ref{sec:Populations} the observed population (see App.~\ref{sec:data}) is characterised and accordingly split into smaller samples. Those subsamples show unexpected tight linear relations, see Sect.~\ref{sec:Results} (additional details can be found in Appendices~\ref{sec:Uncertainties}, \ref{sec:OtherGroups}, and \ref{sec:Posterior}). Possible origins of these relations are discussed in Sect.~\ref{sec:Discussions} leading to the conclusions in Sect.~\ref{sec:Conclusions}.

%--------------------------------------------------------------------
\section{Observed Populations}\label{sec:Populations}
This paper uses the 86 coalescing events reported in the GWTCs with determined masses -- see Table~\ref{tab:data} taken from Table III in \citet[][providing the 11 systems, first observed]{GWTC1}\footnote{The total binary mass is calculated via $M=M_\mathrm{f}+E_\mathrm{rad}\,c^{-2}$.}, Table VI in \citet[][providing 39 new systems]{GWTC2}, and Table IV in \citet[][providing 36 new systems]{GWTC3}. The data includes information on several parameters. Those include five masses (the chirp mass, $\mathcal{M}$, the total binary mass, $M$, the two component masses, $m_1$ and $m_2$, and the final remnant mass, $M_\mathrm{f}$), the effective spin, the luminosity distance, the redshift, and the sky localisation given in all the three catalogues.

Some parameter combinations show strong correlations for a single event, thus these parameters cannot be clearly disentangled. This makes it difficult to use them as independent parameters in population studies. Therefore, the best uncorrelated parameter combinations should be used, which in the case of the mass parameters are the chirp mass and the total mass.

\subsection{The main population}
There is a big and prominent main population in the chirp mass vs. total mass plane. %It follows a linear trend between those two masses and is limited on one side by the maximum mass ratio of 1.
It is limited on one side by the maximum mass ratio\footnote{Gravitational wave observations cannot distinguish the formation order, hence the mass ratio is always defined to be the less massive component devided by the more massive one, which lead to the maximum mass ratio.} of 1 and gets diluted towards smaller mass ratios. Hence, this population is characterized by mass ratios relatively close to 1.

\begin{table}
 \centering
 \caption{\label{tab:members}The members of the two outstanding populations, cf. Fig.~\ref{fig:masses}. They are sorted by increasing total binary mass. The two systems separated in the second group could either belong to that group or the main population.}
 \begin{tabular}{cc}
  \hline
  group~1 & group~2\\
  \hline
  \begin{tabular}{c}
   GW191219\_163120\\
   GW200322\_091133\\
   GW200208\_222617\\
   GW191127\_050227\\
   GW190929\_012149\\
   \\
   \\
  \end{tabular}&
  \begin{tabular}{c}
   GW190814\phantom{\_000000}\\
   GW200210\_092254\\
   GW191113\_071753\\
   GW190412\phantom{\_000000}\\
   GW200306\_093714\\
  \hline
   GW191204\_110529\\
   GW170104\phantom{\_000000}\\
  \end{tabular}\\
%  GW191219\_163120 & GW190814\phantom{\_000000}\\
%  GW200322\_091133 & GW200210\_092254\\
%  GW200208\_222617 & GW191113\_071753\\
%  GW191127\_050227 & GW190412\phantom{\_000000}\\
%  GW190929\_012149 & GW200306\_093714\\
%  \hline
%   & GW191204\_110529\\
%   & GW170104\phantom{\_000000}\\
  \hline
 \end{tabular}
\end{table}

%For mass ratios most distant from 1 while having a total mass above $\unit{20}{\Msun}$ there appear two subpopulations, which will be explained in more detail in the following subsections and are summarized in Table~\ref{tab:members}.
In this study we will focus on the outstanding systems which are most distant from the limiting case of equal masses in the progenitor binary, $dm$, see App.~\ref{sec:data} for the exact definition and its Table~\ref{tab:data} for the values assigned to the events. Starting from the most distant systems we assign systems which do not belong to the main population. While a by eye inspection of the data reveals a quick selection into the populations, see Table~\ref{tab:members}, we present a more quantitative selection in the following subsections by making use of the distance to the limiting boundary of equal masses.

\subsection{Subpopulation 1}
This population contains the most extreme mass ratio system GW191219\_163120 \citep{GWTC3}. %The other members of this group are the event GW190929\_012149 reported in \citet{GWTC2} and the events GW191127\_050227, GW200208\_222617, and GW200322\_091133 reported in \citet{GWTC3}.
The first gap in the distance to the limiting border appears at about $\unit{5-7}{\Msun}$. Two further events, namely GW200322\_091133 and GW200208\_222617 \citep[both reported in][]{GWTC3}, are above this gap. With three systems this population is under represented to constrain more characteristics of this group in a significant way. The second gap in the distance appears at about $\unit{4.0-4.5}{\Msun}$. There are four systems between the two gaps. Those are GW200210\_092254 \citep{GWTC3}, GW190814 \citep{GW190814,GWTC2}, GW190929\_012149 \citep{GWTC2}, and GW191127\_050227 \citep{GWTC3}.

The event GW190929\_012149 had the most massive progenitor system among this group. Because this group contain mainly systems with a mass ratio clearly different from 1, they might suffer from additional uncertainties in the wave form templates, which arise for extremer mass ratios. Hence, they are quoted with large errorbars or in the case of GW191219\_163120 pointed out to may contain additional uncertainties \citep{GWTC3}.

A pure cut according to the distance to the limiting border would suggest this group to consist of seven events with $dm>\unit{4.5}{\Msun}$. The two events GW200210\_092254 and GW190814 appear to be special in this group because of their lower binary mass. Additionally, both of them have a more massive component with a mass $<\unit{25}{\Msun}$ and a less massive component, where it is unclear whether this component is a neutron star or a black hole. Therefore, we do not assign these two systems to belong to group~1. There are 5 remaining events (out of the 86), which represent the first subpopulation investigated in this paper as group~1.

%Tbc.

\subsection{Subpopulation 2}
%This population contains two systems which are believed to have a neutron star and a black hole component.
This population is based on the two systems, which got excluded from group~1. Those are GW190814 \citep{GW190814,GWTC2} and GW200210\_092254 \citep{GWTC3}. It should be noted, that those are the merging systems, where the nature of the less massive component is not fully known. They are suggested to be the most massive known neutron stars with masses of $2.59^{+0.08}_{-0.09}$ and $\unit{2.83^{+0.47}_{-0.42}}{\Msun}$, respectively. This group gets additional members when allowing for $\unit{3}{\Msun}<dm<\unit{4}{\Msun}$. Those new members are the double black hole systems GW191113\_071753 \citep{GWTC3} and GW190412 \citep{GW190412,GWTC2}.

%Beside the five mentioned members there could be two more, which are located in the overlapping region of this group and the main population. Those are GW170104 \citep{aaa+17,GWTC1} and GW191204\_110529 \citep{GWTC3}.
Beside those four systems, there could be more members of this group which are close to or overlapping with the main population. Those are GW200306\_093714 \citep{GWTC3}, GW170104 \citep{aaa+17,GWTC1}, and GW191204\_110529 \citep{GWTC3}. These systems are located closely to a relation, which is established by the four systems, which belong clearly to this group, see Sect.~\ref{sec:Results}. To get a more stable basis of this second group we assign GW200306\_093714 to be a permanent member of group~2. When we talk about all seven events, we will refer to group~2+. Whether or not those systems are included in this group will only cause changes within the uncertainties, see Sect.~\ref{sec:Results} and Table~\ref{tab:fits}.

%Tbc.
\bigskip
We refer the reader to App.~\ref{sec:OtherGroups} for alternative group assignments and the resulting changes for the relations shown in Sect.~\ref{sec:Results}.

%--------------------------------------------------------------------
\section{Results}\label{sec:Results}
\begin{figure*}
 \centering
 \includegraphics[width=\textwidth]{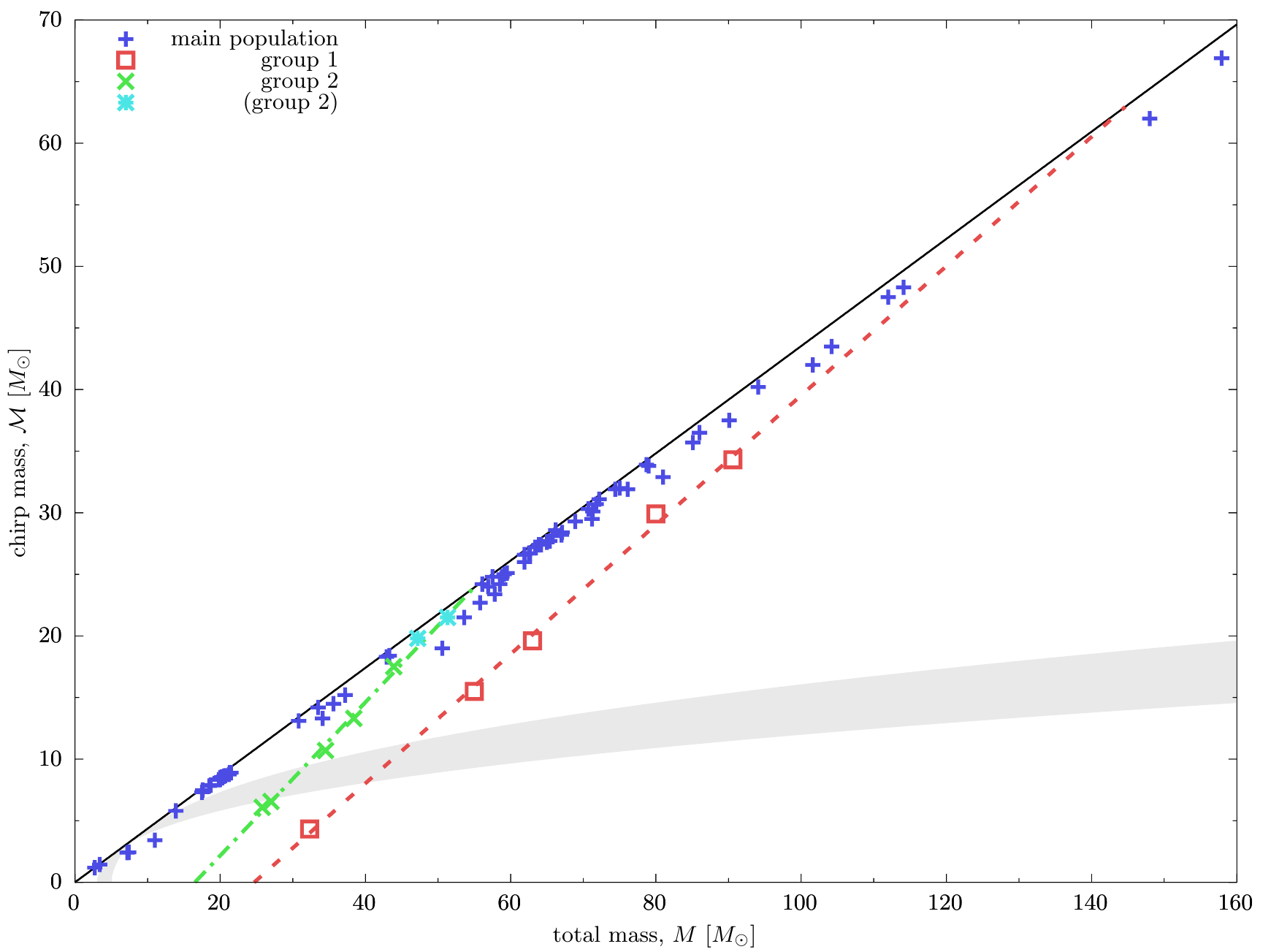}
 \caption{\label{fig:masses}The total and chirp masses of the observed merging events. The colours and point types indicate the group or main population, see key in the top left. The red/dashed and green/dot-dashed lines are fits to the groups~1 and 2, respectively. The black line indicates a mass ratio of 1, hence the space above it is forbidden by definition. The coloured lines are fits to the outstanding groups. The gray area shows, where one component mass is between $3$ and $\unit{5}{\Msun}$, hence within the first mass gap between neutron stars and black holes. The uncertainties are omitted for clarity, but can be found in Fig.~\ref{fig:masses2} in the appendix.}
\end{figure*}
\begin{table}
 \centering
 \caption{\label{tab:fits}Results of fitting linear functions to the values unweighted with $\sigma_\mathcal{M}=\unit{1}{\Msun}$ and a known binary mass for all events. The last four lines show the minimum and maximum total mass of a member of the corresponding group and its uncertainty.}
 \begin{tabular}{cccc}
  \hline
   & group~1 & group~2 & group~2+\\
  \hline
  $a~[M_\odot]$ & $-12.9606$ & $-10.2904$ & $-10.4948$ \\
  $\sigma_{a}~[M_\odot]$ & $\phantom{-0}1.0057$ & $\phantom{-0}1.0553$ & $\phantom{-0}0.7654$ \\
  $b$ & $\phantom{-0}0.5249$ & $\phantom{-0}0.6227$ & $\phantom{-0}0.6300$ \\
  $\sigma_{b}$ & $\phantom{-0}0.0149$ & $\phantom{-0}0.0305$ & $\phantom{-0}0.0194$ \\
  red. $\chi^2$ & $\phantom{-0}0.4578$ & $\phantom{-0}0.2175$ & $\phantom{-0}0.2185$ \\
  \hline
  $M_\mathrm{min}=-\frac{a}{b}~[M_\odot]$ & $\phantom{-}24.6941$ & $\phantom{-}16.5267$ & $\phantom{-}16.6573$ \\
  $\sigma_{M_\mathrm{min}}~[M_\odot]$ & $\phantom{-0}1.2634$ & $\phantom{-0}0.9154$ & $\phantom{-0}0.7244$ \\
  $M_\mathrm{max}=\frac{-a}{b-0.5^{1.2}}~[M_\odot]$ & $144.6975$ & $\phantom{-}54.9178$ & $\phantom{-}53.8847$ \\
  $\sigma_{M_\mathrm{max}}~[M_\odot]$ & $\phantom{-}13.8512$ & $\phantom{-0}3.5943$ & $\phantom{-0}1.8010$ \\
  \hline
 \end{tabular}
\end{table}

In the main part of the paper we will analyse the reported data 
%with a Frequentist approach 
based on the median values (presented as the most reliable ones in the GWTCs), but an analysis taking the full posteriors into account can be found in App.~\ref{sec:Posterior}. At this point it cannot be excluded that the relations presented in the following may be caused or at least supported by biases introduced in the Bayesian methods used to derive the values in the GWTCs. Furthermore, we will focus on the most likely values of the events purely. A consideration of the uncertainties can be found in App.~\ref{sec:Uncertainties}.

The observed systems are shown in the space of total binary mass vs. chirp mass in Fig.~\ref{fig:masses}. Even without marking the groups there, this plot clearly displays outstanding systems, which are furthest away from the limiting line of a mass ratio of one. Hence, it is a natural choice to have a closer look for the systems which are separated from the main population. Additionally, those systems should be least effected by clipping effects caused by the limitation of the mass ratio to not go beyond one. It is evident that these outstanding systems tend to split themselves into two groups (see Table~\ref{tab:members}), marked by colour and symbols. Additionally, there are linear fits for the groups shown, which use
\begin{equation}
 \label{eq:linear}
 \mathcal{M}=a+b\,M.
\end{equation}
This simple linear relation in the parameter space of the total binary mass vs. chirp mass is a non-trivial relation in any other space with mass parameters. For example it translates into a sixth order equation of the individual component masses, see Fig.~\ref{fig:masses4}.

The results of the fits are summarized in Table~\ref{tab:fits}, which uses the values quoted as most probable only (other approaches can be found in the appendix). The relations shown here are very tight, cf. very small reduced $\chi^2$ values. They are much tighter as one would expect from the given measurement uncertainties, see Fig.~\ref{fig:masses2}. The small uncertainties of the fitted parameters emphasise again that the relations are much tighter than expected if those have an origin in the random selection which systems got observed by us from the full population out in the Universe. Because the reduced $\chi^2$ scales with the inverse of the variance one can use the values given in Table~\ref{tab:fits} to calculate a typical uncertainty of the masses required from the observations to confirm or disprove the relations. For a reduced $\chi^2$ of about 1 the required typical observational uncertainty of the chrip masses would be $\sigma_\mathcal{M}\approx\unit{0.68}{\Msun}$ and $\sigma_\mathcal{M}\approx\unit{0.47}{\Msun}$ for the groups~1 and 2, respectively. When taking the uncertainty on the binary mass into account as well even lower observational uncertainties are required. Usually, the chirp mass can be determined with a lower uncertainty than the binary mass. Keeping this in mind, we would need typical uncertainties on the chirp mass $\sigma_\mathcal{M}\lesssim\unit{0.1}{\Msun}$ and the binary mass $\sigma_{M}\lesssim\unit{1}{\Msun}$. The differences in the values for group~2 and when adding the two overlapping systems to group~2+ are indistinguishable within the uncertainties.

It is remarkable, that the relations span over the first mass gap between neutron stars and black holes. This indicates that the underlying reason does either not depend on the compact object types or the less massive components in GW190814, GW200210\_092254, and GW191219\_163120 might be black holes instead of neutron stars.

The relations are limited by physical bounds to a certain range in the total binary mass. Those bounds are the requirement of masses to be positive and the mass ratio to be limited by $1.0$. Hence, a minimum and maximum total binary mass can be calculated for each relation. For both groups the minimum mass is above the maximum binary mass of double neutron star systems. Hence, no double neutron star system can contribute to any of the groups. The maximum mass gives the region where the groups would overlap with the main population. The two potential additional candidates of group~2 are close to the maximum mass allowed for that group. Including the two systems to group~2+ does not change the fit significantly. Therefore, it is an open and probably unsolvable question whether the two systems GW170104 and GW191204\_110529 belong to that group or not. The group~1 would have the overlap with the main population in the second mass gap which is believed to emerge from pair instability, hence there is a lack of observed systems in the overlapping region of group~1. There are a few systems which are close to the relation near the overlapping region. Those are the six systems with the largest estimated masses ($M>\unit{100}{\Msun}$): GW190706\_222641, GW190519\_153544, GW191109\_010717, GW190602\_175927, GW200220\_061928, and GW190521. Any of those events is further away from the fit to group~1 than any member of group~1. Additionally, the four events with $M<\unit{120}{\Msun}$ are all above the relation and the two systems with $M>\unit{140}{\Msun}$ are both below the relation. Thus, it is much less likely that these six systems could belong to group~1 than the two additional systems which may belong to group~2. More details on the results for different group assignments can be found in App.~\ref{sec:OtherGroups}.

If those groups represent very different systems than all the other observations one may expects that those groups show particularities in other parameters as well. Therefore, we checked other parameter spaces of all combinations of the parameters: total binary mass, chirp mass, primary mass, secondary mass, final remnant mass, effective spin, luminosity distance, redshift, and sky localisation. Beside the relation in the total mass vs. chirp mass plane the events do not show any other common feature.

%Tbc.

%-----------------------------------------------------------------
\section{Discussions}\label{sec:Discussions}
The big question is, where does the relations presented in Sect.~\ref{sec:Results} come from. There could be several sources spanning from instrumental artefacts to physical signatures in the formation of double compact object mergers.

\begin{figure}
 \centering
 \includegraphics[width=\columnwidth]{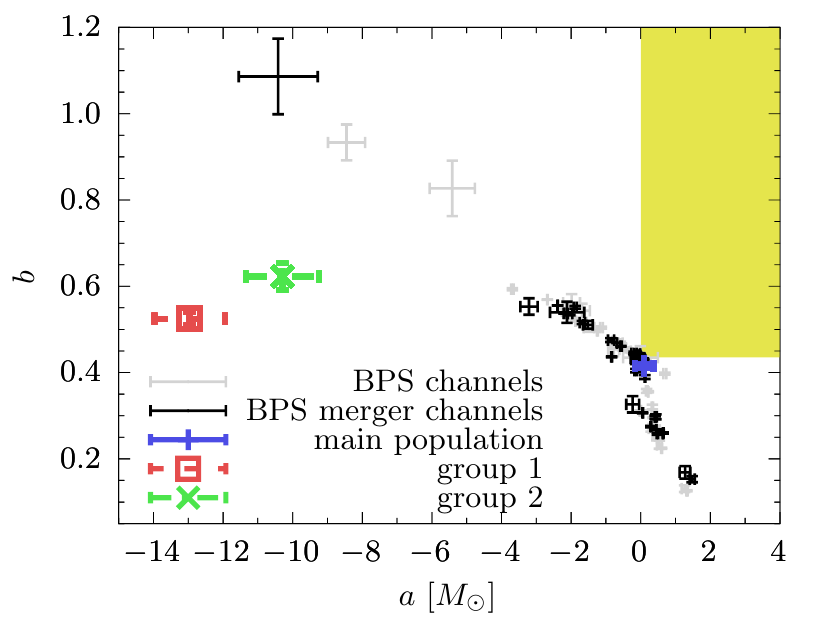}
 \includegraphics[width=\columnwidth]{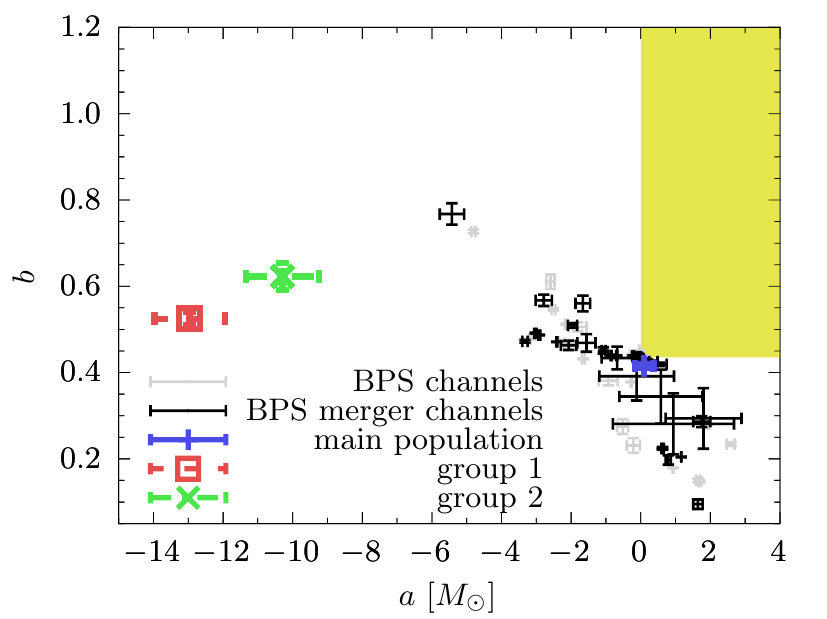}
 \caption{\label{fig:BPS}Offset ($a$) and slope ($b$) parameter of linear fits. The gray and black data points are for different channels in isolated binary evolution at high and low metallicity (top panel: $Z=0.0088$ and bottom panel: $Z=0.0002$), where the black channels lead to double compact object mergers within a Hubble time. The coloured data points represent the fits to the observational data, cf. Table~\ref{tab:fits}. The yellow area is forbidden. Similarly, negative $b$ is disallowed for any negative $a$.}
\end{figure}

A simple linear relation between the total binary and the chirp mass is usually not expected in the formation of a double compact object binary. In the formation of a binary one usually gets relations for different formation channels which show up in the mass ratio instead of the chirp mass. Constant mass ratios would lead to a linear relation between the chirp and total mass, too. But this relation would have $a=0$ and $b=q^{0.6}\,(1+q)^{-1.2}$. The fitting results require an offset clearly different from zero. A possibility of a relation which involves the chirp mass could may arise for systems which suffer strong effects of gravitational wave radiation already during their formation. The stages of evolution where this could happen are very limited because of the requirement of an extremely tight system. Hence, only binaries, where both components have no Hydrogen rich envelope any more are eligible\footnote{Most massive Helium rich envelopes probably be to extended for important gravitational wave radiation in a non contact system, too.}. This condition will be only fulfilled after or directly before the formation of the second formed compact object. But any of such channels would have a common feature, the delay time between the formation and the observable merging event would be shorter than a few $\mega\mathrm{yr}$. Those fast merging systems are less than three percent of the merging population \citep[see e.g.][]{ktl+18}. The members of the subpopulation are more than ten percent of the reported events in the GWTCs, thus seem to be more common than expected from isolated binary evolution. Fig.~\ref{fig:BPS} shows linear fits like Eq.~\eqref{eq:linear} for the different channels of isolated binary evolution presented in \citet{ktl+18}. There is no channel which can match the fits to the two groups, but with the main population. The channels from the simulated data either prefer an offset parameter, $a$, closer to zero, hence a smaller spread in mass ratio or a larger slope parameter (including a larger uncertainty there), which mainly refers to smaller spreads in primary masses. The positive offsets with very low slope correspond to limitations on the secondary mass. A more detailed search for a possible origin in the nature of double compact object mergers may reveal some new insights of so far ignored physics, but is beyond the scope of this paper.

On the other hand, a relation, like reported here, could potentially translate in a relation of signal parameters. The chirp mass should mainly relate to the in-spiral, especially the final orbit(s) with the largest amplitude and therefore best signal-to-noise ratio, in gravitational waves. The total mass is best determined from the ringdown. Hence, the relations here might indicate a relation between the in-spiral and the ringdown in the gravitational wave signal. The observational data of the here mentioned events should be checked for possible relations in their signals, which is beyond the scope of this paper. If there is no support for a relation in the signals itself, the relations might be introduced by the data analysis techniques, e.g. by the choice of priors \citep[the prior influence on masses of double neutron star mergers is recently reported in][]{btv22} or noise reduction. In the case, the signals already show strong relations, an independent check for the different detectors might reveal that the relation is only supported by some of them. This would point to an instrumental artefact, which could be caused by some resonances or damping in a detector. If it turns out that the relations are not related to the physics of the observed systems the relations may serve as an additional tool to identify events, which are destructed by the instruments and/or the data analysis.

When considering the uncertainties of the observations the support for the relations get weaker. This can be interpreted as the relations are of a random nature. This gets more unlikely the more events are part of a subpopulation/relation (see Appendix~\ref{sec:stochastic}), hence we required at least 5 events each. Nevertheless, there is a small chance that the relations are fortuitous. On the other hand, it may indicates that the median values of the derived posteriors are a bad representation of the events. In this case, future reports on the GW observations should report mainly the 90\% credible interval.

There are more potential events during the first three observing runs reported \citep{OGC3,OGC4,GWTC2.1}. None of those events is overlapping with the groups presented here and mainly contribute to the main population. There are two candidates, which have a large $dm$ like the systems investigated, but would not fit onto the relations. This could either weaken the relations or puts more emphasis on the possibility, that the underling assumptions in \citet{GWTC1,GWTC2,GWTC3}, which are altered by \citet{OGC3,OGC4}, may cause the relations. Interestingly, both of the additional events of interest share a similar binary mass and are in the regime to question the existence of a second mass gap, caused by pair instability. Beside having additional events it should be noted, that \citet{OGC4} did not include all the events assigned to groups here, which could be another evidence,that the relations are caused by the analysis used for GWTC3 \citep{GWTC3}, which contains the majority of events in the our groups.

It should be noted that, both events, which suggest a compact object in the mass range of $2.5$ to $\unit{2.9}{\Msun}$, are part of group~2. This could point in the direction that the less massive components of those events are black holes, if the relation would have an origin in the black hole nature, e.g. as primordial black holes. In the regime of primordial black holes less is know, whether there could be any relation, like the ones presented here (Huang, Qing-Guo private communication). Similarly, the formation via dynamical interactions could be another source of the relations. It should be noted, that most dynamical formation scenarios would favour mass ratios closer to one, hence disfavouring slope parameters different from $b=0.5^{1.2}$ and favour offsets close to zero. So, the relations have the potential to become a tool to differentiate between events coming from isolated binary evolution and other origins. On the other hand, an origin in the detectors or the signal analysis may question the derived parameters for those events, especially the high masses of the less massive compact object in the case the less massive components have indeed been neutron stars. In the case of GW190814, it is vastly debated what the nature of the less massive component is \citep[see e.g.][]{bds20,cg20,sl20,rmd+20,vgk20,cccz20,ygb+20,mpwr20,lbb21,tpd+21,zm21,ll21,bdl+21}.

%Tbc.

%-----------------------------------------------------------------
\section{Conclusions}\label{sec:Conclusions}
Here we present for the first time some evidence of relations in the total binary mass vs. chirp mass plane which connect the most outstanding observations of gravitational waves. The nature of those relations is still unclear and should be investigated in future studies. An origin in isolated binary evolution is very unlikely. Other formation pathways like dynamical interactions in clusters or disks of active galactic nuclei, or primordial black holes are expected to be unlikely to produce such a kind of relation, too. The best chance of an origin would be in the signal as the relations between total binary mass and chirp mass may translate to an underlying relation in signal parameters which describe a connection between the spiral-in (including the peak) and the ringdown. A detailed analysis of the signals of all the here mentioned members of the two groups would be required to exclude artefacts introduced by the instruments and/or the data analysis.

The uncertainties quoted for the values derived from observations do weaken the existence of the relations as they weaken the existence of any subpopulations, see Appendices~\ref{sec:Uncertainties}, \ref{sec:OtherGroups}, and \ref{sec:Posterior}. To finally confirm or disprove the relations more observations and/or less uncertain observations are needed. Future observations will show whether there is more support for the relations presented here and whether there are potentially more relations for other subpopulations. In the case the relations remain and become more evident, their origin need to be found. This may increase our understanding of the detectors, some unknown physics, or even serve us a tool to differentiate between different formation scenarios.

%Tbc.

\begin{acknowledgements}
We would like to thank Hailiang Chen and Kaifan Ji for some feedback to this paper. This work is partly supported by Grant No 12090040 and 12090043 of the Natural Science Foundation of China.
%old: 11521303, and 11733008
\end{acknowledgements}

\bibliographystyle{aa+dataset} % style aa.bst
\bibliography{kruckow_refs} % your references Yourfile.bib

\appendix
\section{Data}\label{sec:data}
\begin{table}
 \centering
 \caption{\label{tab:data}The events sorted by distance to equal mass systems in the total mass vs. chirp mass space. The most outstanding events (marked in \textit{italic} and by a $^\ast$) are split into two groups, see Table~\ref{tab:members}.}
 \begin{tabular}{lccc}
  \hline
  event & total mass & chirp mass & distance\\
   & $M~[\Msun]$ & $\mathcal{M}~[\Msun]$ & $dm~[\Msun]$\\
  \hline
  \textit{GW191219\_163120$^\ast$} & $32.3^{+2.2}_{-2.7}$ & $4.32^{+0.12}_{-0.17}$ & $8.930$\\
  \textit{GW200322\_091133$^\ast$} & $55^{+37}_{-27}$ & $15.5^{+15.7}_{-3.7}$ & $7.739$\\
  \textit{GW200208\_222617$^\ast$} & $63^{+100}_{-25}$ & $19.6^{+10.7}_{-5.1}$ & $7.172$\\
  \textit{GW200210\_092254$^\ast$} & $27^{+7.1}_{-4.3}$ & $6.56^{+0.38}_{-0.4}$ & $4.761$\\
  \textit{GW190814$^\ast$} & $25.8^{+1}_{-0.9}$ & $6.09^{+0.06}_{-0.06}$ & $4.713$\\
  \textit{GW190929\_012149$^\ast$} & $90.6^{+21.2}_{-14.1}$ & $34.3^{+8.6}_{-6.5}$ & $4.709$\\
  \textit{GW191127\_050227$^\ast$} & $80^{+39}_{-22}$ & $29.9^{+11.7}_{-9.1}$ & $4.513$\\
  \textit{GW191113\_071753$^\ast$} & $34.5^{+10.5}_{-9.8}$ & $10.7^{+1.1}_{-1}$ & $3.958$\\
  \textit{GW190412$^\ast$} & $38.4^{+3.8}_{-3.7}$ & $13.3^{+0.4}_{-0.3}$ & $3.131$\\
%  \vdots & \vdots & \vdots & \vdots\\
%  \hline
  GW200308\_173609 & $50.6^{+10.9}_{-8.5}$ & $19^{+4.8}_{-2.8}$ & $2.774$\\
  GW200220\_061928 & $148^{+55}_{-33}$ & $62^{+23}_{-15}$ & $2.220$\\
  GW200216\_220804 & $81^{+20}_{-14}$ & $32.9^{+9.3}_{-8.5}$ & $2.161$\\
  GW190706\_222641 & $101.6^{+17.9}_{-13.5}$ & $42^{+8.4}_{-6.2}$ & $2.039$\\
  GW190519\_153544 & $104.2^{+14.5}_{-14.9}$ & $43.5^{+6.8}_{-6.8}$ & $1.701$\\
  GW190513\_205428 & $53.6^{+8.6}_{-5.9}$ & $21.5^{+3.6}_{-1.9}$ & $1.679$\\
  GW190521 & $157.9^{+37.4}_{-20.9}$ & $66.9^{+15.5}_{-9.2}$ & $1.678$\\
  GW200302\_015811 & $57.8^{+9.6}_{-6.9}$ & $23.4^{+4.7}_{-3}$ & $1.613$\\
  GW190620\_030421 & $90.1^{+17.3}_{-12.1}$ & $37.5^{+7.8}_{-5.7}$ & $1.576$\\
  \textit{GW200306\_093714$^\ast$} & $43.9^{+11.8}_{-7.5}$ & $17.5^{+3.5}_{-3}$ & $1.475$\\
%  \vdots & \vdots & \vdots & \vdots\\
  GW190719\_215514 & $55.8^{+16.3}_{-10}$ & $22.7^{+5.9}_{-3.7}$ & $1.456$\\
  GW190828\_065509 & $34.1^{+5.5}_{-4.5}$ & $13.3^{+1.2}_{-0.9}$ & $1.415$\\
  GW190909\_114149 & $71.2^{+54.3}_{-15}$ & $29.5^{+17.5}_{-6.3}$ & $1.368$\\
  GW200105\_162426 & $11^{+1.5}_{-1.4}$ & $3.42^{+0.08}_{-0.08}$ & $1.254$\\
  GW190602\_175927 & $114.1^{+18.5}_{-15.7}$ & $48.3^{+8.6}_{-8}$ & $1.251$\\
  GW170729 & $85.1^{+16.3}_{-11.9}$ & $35.7^{+6.5}_{-4.7}$ & $1.230$\\
  GW190527\_092055 & $58.5^{+27.9}_{-10.6}$ & $24.2^{+11.9}_{-4.4}$ & $1.159$\\
  GW191109\_010717 & $112^{+20}_{-16}$ & $47.5^{+9.6}_{-7.5}$ & $1.147$\\
  GW190413\_134308 & $76.1^{+15.9}_{-10.6}$ & $31.9^{+7.3}_{-4.6}$ & $1.123$\\
  GW190512\_180714 & $35.6^{+3.9}_{-3.4}$ & $14.5^{+1.3}_{-1}$ & $0.913$\\
  GW151012 & $37.2^{+10.4}_{-4.3}$ & $15.2^{+2}_{-1.1}$ & $0.910$\\
  GW200220\_124850 & $67^{+17}_{-12}$ & $28.2^{+7.3}_{-5.1}$ & $0.883$\\
  GW190517\_055101 & $61.9^{+10}_{-9.6}$ & $26^{+4.2}_{-4}$ & $0.865$\\
  GW190503\_185404 & $71.3^{+9.3}_{-8}$ & $30.1^{+4.2}_{-4.1}$ & $0.857$\\
  GW191230\_180458 & $86^{+19}_{-12}$ & $36.5^{+8.2}_{-5.6}$ & $0.856$\\
  \textit{GW170104$^\ast$} & $51.3^{+5.7}_{-4.4}$ & $21.5^{+2.1}_{-1.7}$ & $0.761$\\
%  \vdots & \vdots & \vdots & \vdots\\
  GW190731\_140936 & $67.1^{+15.3}_{-10.2}$ & $28.4^{+6.8}_{-4.5}$ & $0.740$\\
  GW190915\_235702 & $59.5^{+7.5}_{-6.2}$ & $25.1^{+3.1}_{-2.6}$ & $0.732$\\
  GW190630\_185205 & $58.8^{+4.7}_{-4.8}$ & $24.8^{+2.1}_{-2}$ & $0.728$\\
  GW200115\_042309 & $7.4^{+1.8}_{-1.7}$ & $2.43^{+0.05}_{-0.07}$ & $0.725$\\
  GW190413\_052954 & $56.9^{+13.1}_{-8.9}$ & $24^{+5.4}_{-3.7}$ & $0.703$\\
  GW200208\_130117 & $65.4^{+7.8}_{-6.8}$ & $27.7^{+3.6}_{-3.1}$ & $0.703$\\
  GW190701\_203306 & $94.1^{+11.6}_{-9.3}$ & $40.2^{+5.2}_{-4.7}$ & $0.696$\\
  \textit{GW191204\_110529$^\ast$} & $47.2^{+9.2}_{-8}$ & $19.8^{+3.6}_{-3.3}$ & $0.683$\\
%  \vdots & \vdots & \vdots & \vdots\\
  GW170809 & $59.1^{+5.8}_{-4.3}$ & $25^{+2.1}_{-1.6}$ & $0.665$\\
  GW190426\_152155 & $7.2^{+3.5}_{-1.5}$ & $2.41^{+0.08}_{-0.08}$ & $0.664$\\
  GW200219\_094415 & $65^{+12.6}_{-8.2}$ & $27.6^{+5.6}_{-3.8}$ & $0.635$\\
  GW170823 & $68.9^{+10.3}_{-7.4}$ & $29.3^{+4.2}_{-3.2}$ & $0.633$\\
  \hline
 \end{tabular}
\end{table}
\begin{table}
 \centering
 \caption{Table~\ref{tab:data} continued.}
 \begin{tabular}{lccc}
  \hline
  event & total mass & chirp mass & distance\\
   & $M~[\Msun]$ & $\mathcal{M}~[\Msun]$ & $dm~[\Msun]$\\
  \hline
  GW200128\_022011 & $75^{+17}_{-12}$ & $32^{+7.5}_{-5.5}$ & $0.592$\\
  GW190803\_022701 & $62.7^{+11.8}_{-8.4}$ & $26.7^{+5.2}_{-3.8}$ & $0.543$\\
  GW191222\_033537 & $79^{+16}_{-11}$ & $33.8^{+7.1}_{-5}$ & $0.538$\\
  GW190421\_213856 & $71.8^{+12.5}_{-8.6}$ & $30.7^{+5.5}_{-3.9}$ & $0.507$\\
  GW200209\_085452 & $62.6^{+13.9}_{-9.4}$ & $26.7^{+6}_{-4.2}$ & $0.503$\\
  GW190514\_065416 & $64.2^{+16.6}_{-9.6}$ & $27.4^{+6.9}_{-4.3}$ & $0.499$\\
  GW190727\_060333 & $65.8^{+10.9}_{-7.4}$ & $28.1^{+4.9}_{-3.4}$ & $0.496$\\
  GW170818 & $62.5^{+5.3}_{-4.3}$ & $26.7^{+2.1}_{-1.7}$ & $0.463$\\
  GW190521\_074359 & $74.4^{+6.8}_{-4.6}$ & $31.9^{+3.1}_{-2.4}$ & $0.444$\\
  GW200316\_215756 & $21.2^{+7.2}_{-2}$ & $8.75^{+0.62}_{-0.55}$ & $0.438$\\
  GW190424\_180648 & $70.7^{+13.4}_{-9.8}$ & $30.3^{+5.7}_{-4.2}$ & $0.435$\\
  GW151226 & $21.5^{+6.5}_{-1.7}$ & $8.9^{+0.3}_{-0.3}$ & $0.420$\\
  GW191215\_223052 & $43.3^{+5.3}_{-4.3}$ & $18.4^{+2.2}_{-1.7}$ & $0.410$\\
  GW200112\_155838 & $63.9^{+5.7}_{-4.6}$ & $27.4^{+2.6}_{-2.1}$ & $0.380$\\
  GW200129\_065458 & $63.4^{+4.3}_{-3.6}$ & $27.2^{+2.1}_{-2.3}$ & $0.364$\\
  GW200225\_060421 & $33.5^{+3.6}_{-3}$ & $14.2^{+1.5}_{-1.4}$ & $0.350$\\
  GW190408\_181802 & $42.9^{+4.1}_{-2.9}$ & $18.3^{+1.8}_{-1.2}$ & $0.342$\\
  GW190720\_000836 & $21.3^{+4.3}_{-2.3}$ & $8.9^{+0.5}_{-0.8}$ & $0.341$\\
  GW191103\_012549 & $20^{+3.7}_{-1.8}$ & $8.34^{+0.66}_{-0.57}$ & $0.335$\\
  GW191126\_115259 & $20.7^{+3.4}_{-2}$ & $8.65^{+0.95}_{-0.71}$ & $0.330$\\
  GW190910\_112807 & $78.7^{+9.5}_{-9}$ & $33.9^{+4.3}_{-3.9}$ & $0.327$\\
  GW200311\_115853 & $61.9^{+5.3}_{-4.2}$ & $26.6^{+2.4}_{-2}$ & $0.315$\\
  GW190930\_133541 & $20.3^{+9}_{-1.5}$ & $8.5^{+0.5}_{-0.5}$ & $0.308$\\
  GW200224\_222234 & $72.2^{+7.2}_{-5.1}$ & $31.1^{+3.2}_{-2.6}$ & $0.300$\\
  GW190728\_064510 & $20.5^{+4.5}_{-1.3}$ & $8.6^{+0.5}_{-0.3}$ & $0.296$\\
  GW191129\_134029 & $17.5^{+2.4}_{-1.2}$ & $7.31^{+0.43}_{-0.28}$ & $0.282$\\
  GW190708\_232457 & $30.8^{+2.5}_{-1.8}$ & $13.1^{+0.9}_{-0.6}$ & $0.281$\\
  GW191216\_213338 & $19.81^{+2.69}_{-0.94}$ & $8.33^{+0.22}_{-0.19}$ & $0.268$\\
  GW190924\_021846 & $13.9^{+5.1}_{-0.9}$ & $5.8^{+0.2}_{-0.2}$ & $0.230$\\
  GW191204\_171526 & $20.21^{+1.7}_{-0.96}$ & $8.55^{+0.38}_{-0.27}$ & $0.226$\\
  GW170608 & $18.7^{+3.2}_{-0.8}$ & $7.9^{+0.2}_{-0.2}$ & $0.220$\\
  GW191105\_143521 & $18.5^{+2.1}_{-1.3}$ & $7.82^{+0.61}_{-0.45}$ & $0.213$\\
  GW190828\_063405 & $57.5^{+7.5}_{-4.4}$ & $24.8^{+3.3}_{-2}$ & $0.209$\\
  GW170814 & $56.1^{+3.6}_{-2.7}$ & $24.2^{+1.4}_{-1.1}$ & $0.201$\\
  GW150914 & $66.2^{+3.7}_{-3.4}$ & $28.6^{+1.6}_{-1.5}$ & $0.197$\\
  GW190707\_093326 & $20^{+1.9}_{-1.3}$ & $8.5^{+0.6}_{-0.4}$ & $0.188$\\
  GW200202\_154313 & $17.58^{+1.78}_{-0.67}$ & $7.49^{+0.24}_{-0.2}$ & $0.149$\\
  GW190425 & $3.4^{+0.3}_{-0.1}$ & $1.44^{+0.02}_{-0.02}$ & $0.037$\\
  GW170817 & $2.73^{+0.21}_{-0.19}$ & $1.186^{+0.001}_{-0.001}$ & $0.002$\\
  \hline
 \end{tabular}
\end{table}
Table~\ref{tab:data} shows the data used in this paper. it is taken from \citet{GWTC1,GWTC2,GWTC3}. The distance of each data point to the line of an equal mass ratio is calculated via
\begin{equation}
 dm = \frac{0.5^{\frac{6}{5}}\,M-\mathcal{M}}{\sqrt{1+0.5^{\frac{12}{5}}}} .
\end{equation}
The distance cut at $dm\approx\unit{3}{\Msun}$ is arbitrary, but can be motivated by Fig.~\ref{fig:distances} to separate the two local maxima at high $dm$. The system just below this cut, GW200308\_173609, may be part of a third group. But there is not enough evidence to claim a third group there yet. With future observations we might get more groups which follow linear relations in the total mass vs. chirp mass space.

\section{Measurement Uncertainties}\label{sec:Uncertainties}
\begin{figure*}
 \centering
 \includegraphics[width=\textwidth]{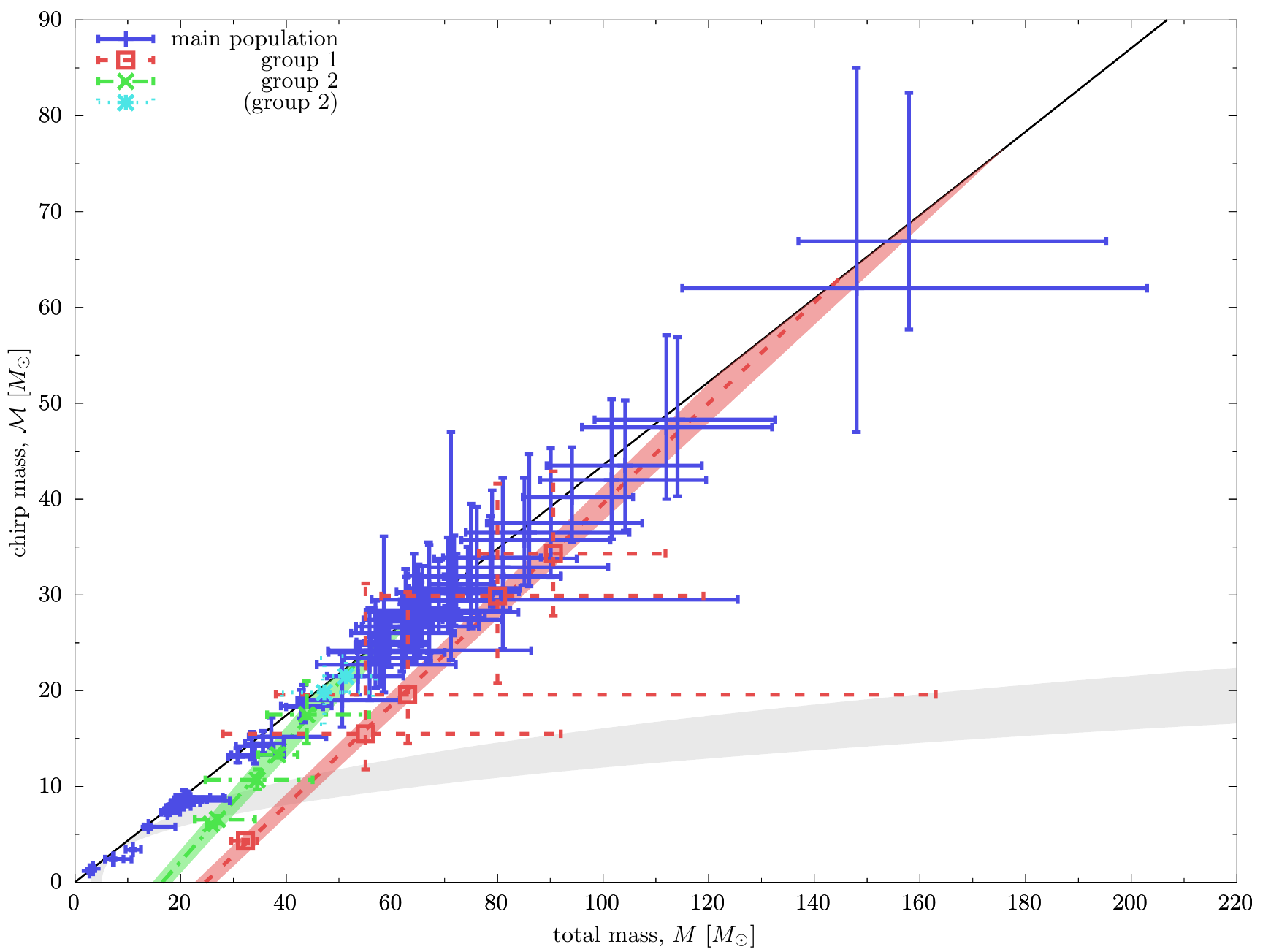}
 \caption{\label{fig:masses2}Similar as Fig.~\ref{fig:masses} but including uncertainties of the data (90\% credible intervals) and the fits ($1\sigma$ uncertainties).}
\end{figure*}
\begin{table}
 \centering
 \caption{\label{tab:fits_weighting}Results of fitting linear functions to the values weighted according to the quoted interval of 90\% credibility.}
 \begin{tabular}{cccc}
  \hline
   & group~1 & group~2 & group~2+\\
  \hline
  $a~[M_\odot]$ & $-12.38612$ & $-8.91177$ & $-9.41245$ \\
  $\sigma_{a}~[M_\odot]$ & $\phantom{-0}0.21107$ & $\phantom{-}0.43845$ & $\phantom{-}0.32948$ \\
  $b$ & $\phantom{-0}0.51718$ & $\phantom{-}0.58109$ & $\phantom{-}0.60018$ \\
  $\sigma_{b}$ & $\phantom{-0}0.00619$ & $\phantom{-}0.01627$ & $\phantom{-}0.01180$ \\
%  red. $\chi^2$ & $\phantom{-0}0.00124$ & $\phantom{-}0.01045$ & $\phantom{-}0.01300$ \\
  red. $\chi^2$ & $\phantom{-0}0.00334$ & $\phantom{-}0.02828$ & $\phantom{-}0.03517$ \\
  \hline
 \end{tabular}
\end{table}

Fig.~\ref{fig:masses2} includes the uncertainties, which are omitted in the main part. Beside the uncertainties of the events the uncertainties of the unweighted fits to the most likely values are shown, too. It should be noted, that the uncertainties of the fits do not represent the uncertainties of the events.

Additionally, we show in Table~\ref{tab:fits_weighting} the fit results when using the uncertainty given for the individual evens for weighting. The quoted 90\% credible interval of each event is transformed into a standard deviation under the assumption of a Gaussian distribution, hence corresponding to $3.28971 \sigma$. We have to caution the reader that this rescaling is not fully correct because of the non Gaussian posterior distributions. %The half interval of the quoted area of 90\% credibility is taken as one $\sigma$ region to calculate the weights here. 
As a consequence of the large uncertainties the values of the reduced $\chi^2$ are significantly smaller than for the unweighted case. It should be noted, that only the stochastic uncertainty should be used for a $\chi^2$ fit, thus the reported values in Table~\ref{tab:fits_weighting} only serve lower limits on the reduced $\chi^2$. With the assumption that all the uncertainties are dominated by stochastic effects, the real reduced $\chi^2$ values should differ by less than a factor for four. In this case, the values outline a too good fit. Thus, there is the possibility that the relations are too tight to be well confirmed. On the hand, those values cannot exclude the relations either. Therefore, it should be mainly interpreted as a first evidence for the relations. More precise observations of events with unequal masses are needed for a final confirmation of the relations.

While the fits for group~1 and 2+ are consistent with the values presented in Table~\ref{tab:fits}, the values for group~2 differ. This is mainly caused by the event GW190814. It has significantly smaller relative uncertainties than all other events (only GW170817 has a tighter constrained chirp mass). Hence, GW190814 dominates a weighted fit to the group~2. In the case of group~2+, the additional two data points partly counter balance this domination.

\section{Other Group Assignments}\label{sec:OtherGroups}
There is a certain arbitrariness in choice of the groups. Therefore, we investigate here the influence of taking other events into account.

\subsection{Alternating Group 1}
\begin{table}
 \centering
 \caption{\label{tab:fits1}Results of fitting linear functions to the unweighted values for different choices of group~1. The quoted uncertainties are one $\sigma$.}
 \begin{tabular}{cccc}
  \hline
  group & $a~[M_\odot]$ & $b$ & red. $\chi^2$\\
  \hline
  1  & $-12.9606\pm1.0057$ & $0.52485\pm0.01494$ & $0.45783$ \\
  \hline
  1a & $-11.3339\pm0.6875$ & $0.49754\pm0.00690$ & $0.63597$ \\
  1b & $-14.0442\pm0.8223$ & $0.54429\pm0.00991$ & $0.53647$ \\
  1c & $-14.1673\pm1.0081$ & $0.54735\pm0.01274$ & $0.68103$ \\
  1+ & $-11.1983\pm1.3980$ & $0.50723\pm0.01360$ & $2.65060$ \\
  \hline
 \end{tabular}
 \begin{flushleft}
  \emph{group~1}: as defined in Sect.~\ref{sec:Populations}\\
  \emph{group~1a}: group~1 plus events with $M>\unit{140}{\Msun}$ (GW200220\_061928 and GW190521)\\
  \emph{group~1b}: group~1 plus events with $\unit{110}{\Msun}<M<\unit{115}{\Msun}$ (GW190602\_175927 and GW191109\_010717)\\
  \emph{group~1c}: group~1 plus events with $\unit{100}{\Msun}<M<\unit{105}{\Msun}$ (GW190706\_222641 and GW190519\_153544)\\
  \emph{group~1+}: group~1 plus events with $M>\unit{100}{\Msun}$
 \end{flushleft}
\end{table}

Here we allow additional members for group~1. Possible candidates are the six systems with binary masses above $\unit{100}{\Msun}$. Those are GW200220\_061928 \citep{GWTC3}, GW190706\_222641 \citep{GWTC2}, GW190519\_153544 \citep{GWTC2}, GW190521 \citep{GWTC2}, GW190602\_175927 \citep{GWTC2}, and GW191109\_010717 \citep{GWTC3}.

Table~\ref{tab:fits1} summarizes the results of fits using different definitions of group~1. The inclusion of additional events results in larger changes for the fit parameters especially for the slope. All alternative choices result in an increase of the reduced $\chi^2$, which indicates that an inclusion of more systems into group~1 is disfavoured if the group follows a linear relation. It should be noted, that including all events with a binary mass above $\unit{100}{\Msun}$ is the worst choice. This is caused by the fact that the six additional systems could not fall onto a linear relation with an offset parameter significantly smaller than zero.

\subsection{Alternating Group 2}
\begin{table}
 \centering
 \caption{\label{tab:fits2}Results of fitting linear functions to the unweighted values for different choices of group~2. The quoted uncertainties are one $\sigma$.}
 \begin{tabular}{cccc}
  \hline
  group & $a~[M_\odot]$ & $b$ & red. $\chi^2$\\
  \hline
  2   & $-10.2904\pm1.0553$ & $0.62265\pm0.03050$ & $0.21749$ \\
  2+  & $-10.4948\pm0.7654$ & $0.63004\pm0.01944$ & $0.21849$ \\
  \hline
  2-  & $-\phantom{0}8.7696\pm0.6702$ & $0.57063\pm0.02104$ & $0.04839$ \\
  2++ & $-10.0960\pm0.6765$ & $0.61819\pm0.01618$ & $0.22377$ \\
  2+++& $-\phantom{0}9.4952\pm0.6630$ & $0.60055\pm0.01463$ & $0.27826$ \\
  \hline
 \end{tabular}
\end{table}

\subsubsection{Group 2-}
\label{sec:group2m}
Here we use only the four systems: GW200210\_092254, GW190814, GW191113\_071753, and GW190412, which satisfy $dm>\unit{3}{\Msun}$ and do not belong to group~1. The results for this smaller group are shown in Table~\ref{tab:fits2}. The values differ when GW200306\_093714 is not included into this group. The values of the reduced group are similar to the values obtained from the weighted fit, where GW190814 is dominating. We have to caution the reader that for group~2- the number of free parameters is equal to the number of degrees of freedom, i.e. half the number of data points, and therefore this fit have a lower support by the data.

\subsubsection{Group 2++}
Here we allow another additional member with a mass ratio very close to 1. This is the event GW170814 \citep{GWTC1}. In this case, there is only a small change similarly to the change from group~2 to group~2+.

For being more generous we include GW190828\_063405 and GW190413\_052954 to arrive at group~2+++. The results of this group look like a medium of group~2/2+ and group~2-. So, the most extreme change for the relation of group~2 are between group~2- and group~2+.

\bigskip

Nevertheless, all the applied alternations to fit the most likely values of chirp mass and binary mass in Appendices~\ref{sec:Uncertainties} and \ref{sec:OtherGroups} remain in a similar area of the two free parameters. This part of the parameter space $a<-8$ and $b<0.65$ seems to be exclusive for the groups derived from the observations, cf. Fig~\ref{fig:BPS}.
The results of the different group assignments in Tables~\ref{tab:fits1} and \ref{tab:fits2} can be used to estimate a systematic uncertainty caused by the assignments. For group~1 the mean and standard deviation of the different assignments are $a [\Msun]=-12.7409\pm1.2760$ and $b=0.5243\pm0.0197$. Similarly one gets for group~2 $a [\Msun]=-9.8292\pm0.6263$ and $b=0.6084\pm0.0212$. This systematic uncertainty is a bit larger but very similar to the statistical one reported in the tables. While for group~1 the original assignment in the main part is closest to the mean, in the case of group~2 it is the group assignment with the most events (2+++), which is closest.

\subsection{Clustering of Linear Relations}
\begin{figure}
 \centering
 \includegraphics[width=\columnwidth]{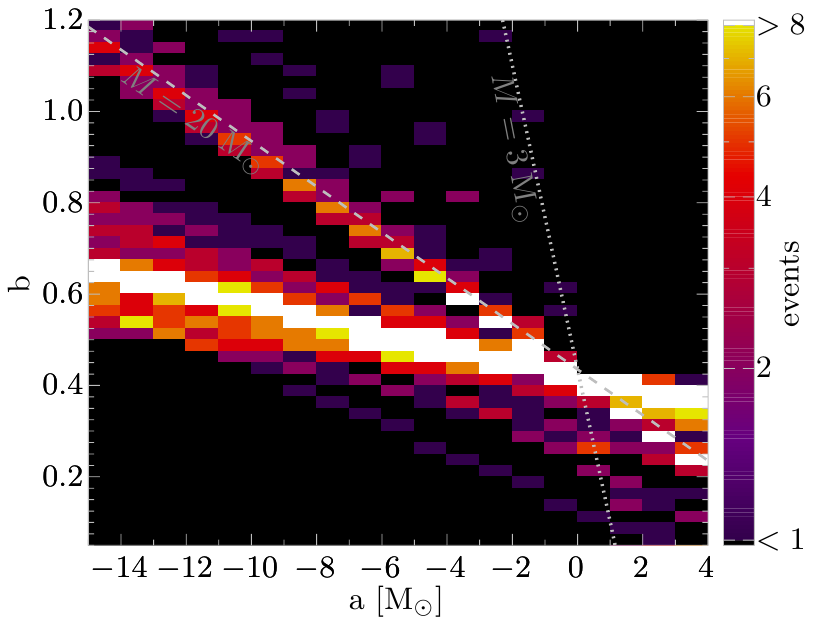}
 \caption{\label{fig:linear1}Heat map in the space of the linear fit parameters, for more details see text. The gray dotted/dashed lines use a mass ratio of 1.}
\end{figure}
To be more independent of the group assignment the problem can be reversed. By considering the function
\begin{equation}
  b(a) = \frac{\mathcal{M}-a}{M},
\end{equation}
which is the same as Eq.~\eqref{eq:linear}. Here each data point of an GW event describes a linear function in the parameter space of the two variables $a$ and $b$. This methodology is similar to a Hough transform to detect lines but with the back mapping to the parameters we used for the fitting. By sampling those functions one can create a heat map, see Fig.~\ref{fig:linear1} where the samples are taken as the mean of each bin in $a$. For the resolution of the heat map we choose values similar to the typical one $\sigma$ uncertainties we got earlier: $\delta a=\unit{1}{\Msun}$ and $\delta b=0.025$. In Fig.~\ref{fig:linear1} is a big covering caused by the main population. Additionally, there is a pronounced secondary linear relation with the gray dashed line named $M=\unit{20}{\Msun}$. This gray line indicates a fictive event with this binary mass and a mass ratio of 1 and therefore $\mathcal{M}\approx\unit{8.7}{\Msun}$. This corresponds to the first peak in the chirp mass distribution reported in \citet{aaa+21}, which is followed by an under density when going to higher masses before getting to the rest of the more massive double black hole mergers. For a mass ratio of 1 all the relations in the $ab$ space move through $a=0$ and $b=0.5^{1.2}$ with a slope of $-M^{-1}$. The two double neutron star mergers are well visible as well, see the second gray dashed line with $M=\unit{3}{\Msun}$. Unfortunately, our groups are mainly hidden behind the main population in Fig.~\ref{fig:linear1}.

\begin{figure}
 \centering
 \includegraphics[width=\columnwidth]{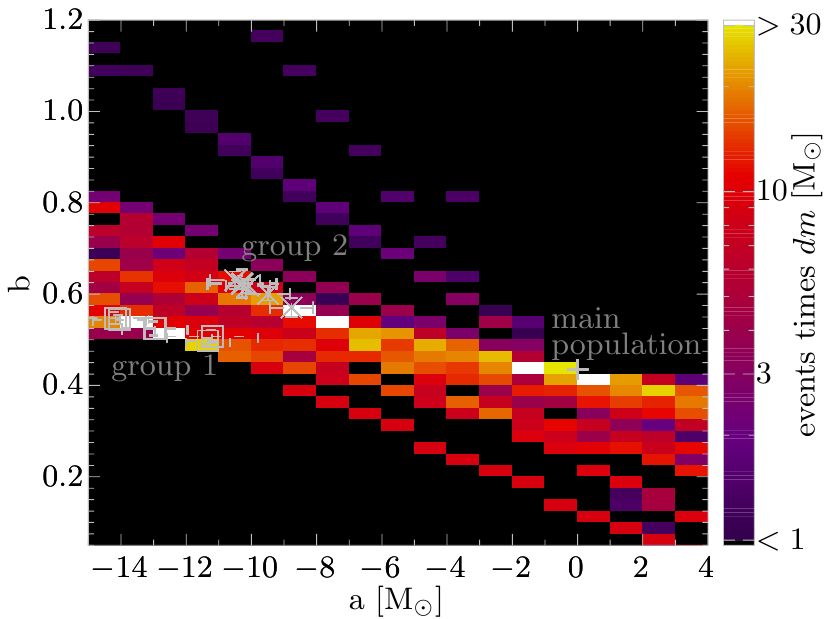}
 \caption{\label{fig:linear2}Similar to Fig.~\ref{fig:linear1} but for each data point the $dm$ is summed to create a colour scale to emphasise the events with large $dm$. The results for the group fits shown in Tables~\ref{tab:fits1} and \ref{tab:fits2} are indicated in gray. For the main population we use the line of equal mass ratio, being a point in this plot.}
\end{figure}
In this paper we are interested in the systems with mass ratios different from 1, hence we emphasise those events by making use of the $dm$ when creating in Fig.~\ref{fig:linear2}, which is otherwise the same as Fig.~\ref{fig:linear1}. The double neutron star mergers and the least massive black hole mergers are below the colour scale threshold and therefore not visible in the Fig.~\ref{fig:linear2}. The main population gets a lower importance because its strength in number is partly suppressed by their usually small $dm$. The main intersection across all the masses of the main population is still visible around $a=0$ and $b=0.5^{1.2}$ (marked by the gray plus). Beside the main population Fig.~\ref{fig:linear2} shows two additional local maxima at $a\in[-14,-12]$ and $b\in[0.50,0.55]$ (matching our group~1, marked by gray squares) and $a\in[-9,-7]$ and $b\in[0.525,0.575]$ (matching our group~2-, marked by gray crosses). For the maximum corresponding to the second group there is still some influence of the main population penetrating there and therefore shifting the maxima to slightly toward $a=0$ and $b=0.5^{1.2}$ as a group independent from the main population would prefer. It should be noted that the local maxima visible in Fig.~\ref{fig:linear2} are independent on our group assignments, while matching the fit results we obtained for our groups. This shows that our group assignments do not have a strong bias as one might expect from a by eye assignment.

\section{Other Mass Spaces}\label{sec:OtherMasses}
\begin{figure}
 \centering
 \includegraphics[width=\columnwidth]{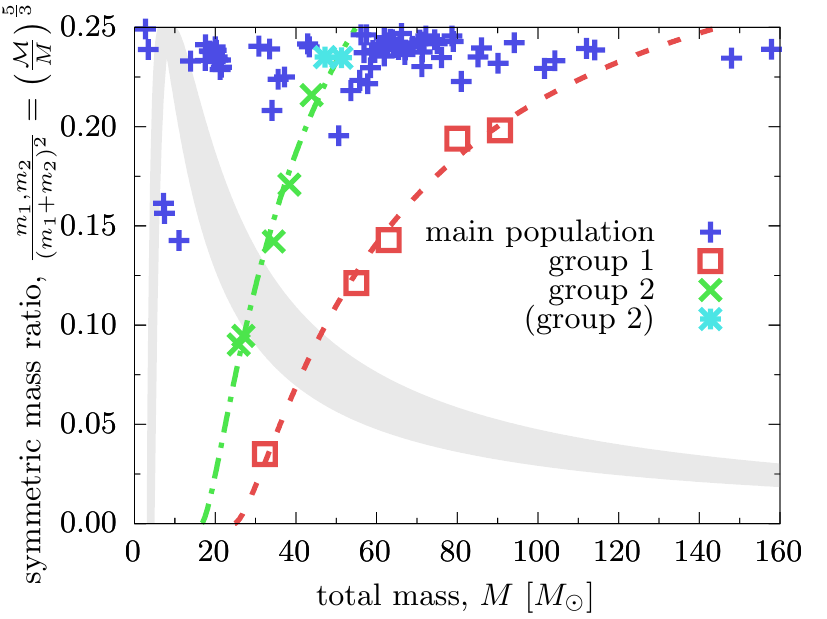}
 \caption{\label{fig:masses3}Similar as Fig.~\ref{fig:masses} but showing the symmetric mass ratio instead of the chirp mass. Again uncertainties are omitted for clarity.}
\end{figure}
\begin{figure}
 \centering
 \includegraphics[width=\columnwidth]{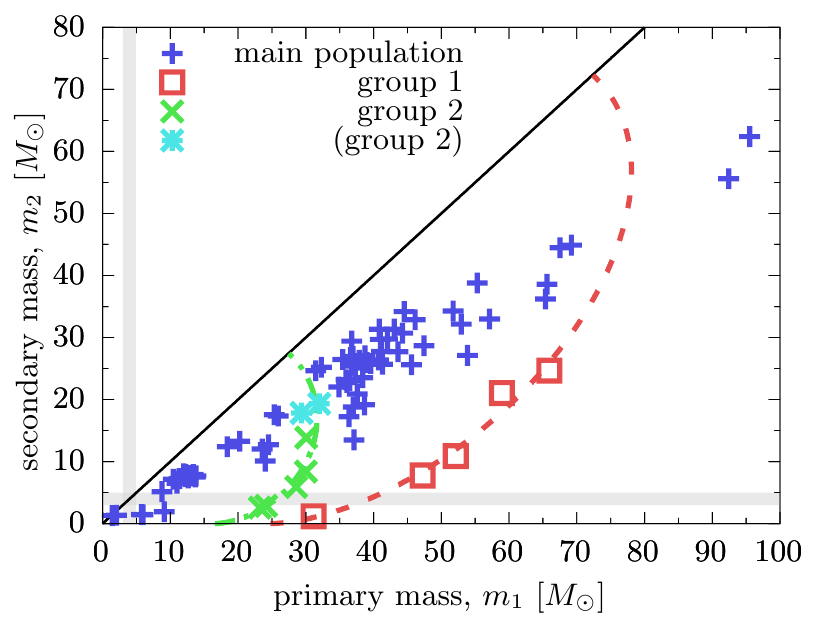}
 \caption{\label{fig:masses4}Similar as Fig.~\ref{fig:masses} but showing two component masses. Again uncertainties are omitted for clarity.}
\end{figure}
Fig.~\ref{fig:masses3} shows the data in a different space using the symmetric mass ratio instead of the chirp mass, but calculated from the chirp and total mass values. In this space the simple linear relations are curved. Fig.~\ref{fig:masses4} shows the plane with the two component masses, which is more preferred from the theoretical studies on GW events. Unfortunately, those two masses are highly correlated when deriving them from observational signals. Hence, they serve bad as independent variables in any analysis of observational data. It should be noted, that the here shown masses are calculated from the total and chirp masses via 
\begin{equation}
 m_{1,2} = \left[\frac{1}{2}\pm\sqrt{\frac{1}{4}-\left(\frac{\mathcal{M}}{M}\right)^{\frac{5}{3}}}\right]\,M .
\end{equation}
Hence, they may differ a bit from the quoted individual masses in the GWTCs because of the strong correlation between the individual masses and the method used to derive the values in the GWTCs.

The other parameter spaces do not show any other relation which would be simpler than the relation derived from the chirp mass vs. total-mass plane.

\section{Analysis of the Full Posterior Data}\label{sec:Posterior}
Here, we present the results of an analysis to confirm the relations by making use of the published posterior distributions. To make use of the full posterior distributions the data need to be analysed by a Monte Carlo approach or a Bayesian analysis. We have to caution the reader, that a Bayesian analysis introduces biases by construction\footnote{E.g. any difference between the true, but unknown, and the assumed prior and/or likelihood functions will introduce a bias.}. The main attempt of such an analysis is to compensate for known biases. But it can not be excluded that new biases are introduced to the data, especially against unknown aspects, nor that all known biases are fully compensated \citep[For more details on possible issues of Bayesian tools we refer the reader to the recent summary by][and references therein]{rtl22}. For better comparison we follow the same approach as used in \citet[][and references therein]{aaa+21} to analyse the population of systems responsible for the observed events. The cut at $\mathrm{FAR}=\unit{0.25}{\mathrm{yr}^{-1}}$ removes some of the systems of interest, hence we need to relax this to include all the systems shown in Table~\ref{tab:data}.

%\subsection{Posterior Data}
For all events reported in GWTC-1 \citep{GWTC1,GWTC1data} we calculate the source masses by assuming $H_0=\unit{67.9}{\kilo\meter\,\second^{-1}\,\mega\mathrm{pc}^{-1}}$ and $\Omega_\mathrm{m}=0.3065$ in a flat cold dark matter cosmology \citep{aaa+21}. In the case of GW170817 we use the low spin posterior, which is favoured by \citet{aaa+17c}. For all events reported in GWTC-2 \citep{GWTC2,GWTC2data} we use the \texttt{PublicationSamples} posteriors in the \texttt{*comoving.h5} files. For all events reported in GWTC-3 \citep{GWTC3,GWTC3data} we use the \texttt{C01:Mixed} posteriors\footnote{\citet{GWTC3} does not state the value of the likelihood cut made in Table IV, hence we include all published sample points.} in the \texttt{*cosmo.h5} files. We used the published files in their version from January 2022.

Additionally, we need to caution the reader that some systems show differences in the values quoted from the catalogue papers and the values one gets from the published posteriors. In the case of the data from GWTC-1 the small differences are mainly caused by taking the fourth digit of $\Omega_\mathrm{m}$ into account, which is not done in \citet{GWTC1} and causes a small difference in the source masses. In the case of GWTC-2 data there are several posterior samples given and for some systems another sample coincides with the data given in the paper better, e.g. for GW190519\_153544 the \texttt{PrecessingSpinIMRHM} posterior matched the published values better. Additionally, there are some systems, where the values differ between different versions of \citet{GWTC2}. All those small differences will cause small differences between the following analysis and the results presented in the main part.

\begin{figure}
 \centering
 \includegraphics[width=\columnwidth]{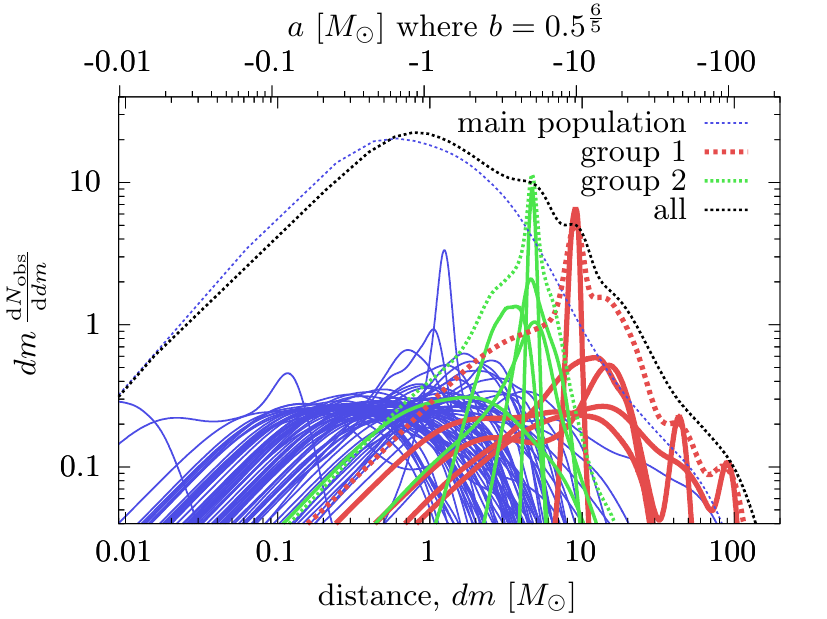}
 \caption{\label{fig:distances}Similar as the top panel of Fig.~2 in \citet{aaa+21} but showing distance, $dm$, instead of the chirp mass, $\mathcal{M}$. The solid lines show the posteriors of the individual events, while the dashed lines are the overall distribution of the 86 events (black) or the subpopulations. The events of the main population are shown in blue/thin, while groups~1 and 2 are red/thick and green/medium, respectively.}
\end{figure}
Fig.~\ref{fig:distances} shows the posterior distributions of the 86 events. It should be noted, that the linear decline on the left hand side of wide curves (spanning more than 2 orders of magnitude) are caused by resolution issues and therefore even partly cut of by the choice of the lower boundary in $dm$. The overall distribution shows one main peak caused by the main population. Additionally, it shows two over densities on the right tail. Those are mainly caused by the members of groups~1 and 2. The events GW191219\_163120 and GW200322\_091133 do show a bimodal distribution as mentioned in \citet{GWTC3}. The event GW200306\_093714 has a lower $dm$ as some other events from the main population (see Table~\ref{tab:data}), which makes it a questionable member of group~2 (see App.~\ref{sec:group2m}) in Fig.~\ref{fig:distances}, too. Fig.~\ref{fig:distances} confirms the dominance of group~1 members for $dm\gtrsim7.0$ and the dominance of group~2 members for $4.0\lesssim dm\lesssim5.5$ as seen in Table~\ref{tab:data}.

\begin{figure}
 \centering
 \includegraphics[width=\columnwidth]{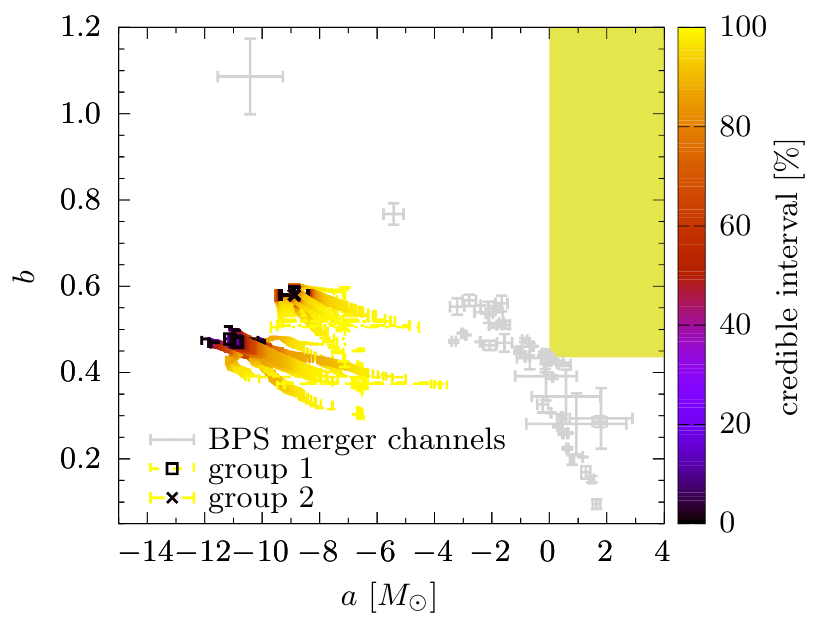}
 \caption{\label{fig:CredibileLevel}Similar to Fig.~\ref{fig:BPS}. The BPS merging channels (gray) include both high and low metallicity. The coloured data points are fits to different credible intervals of the posteriors of the events of group~1 (squares, the lower sequence) and group~2 (crosses, the upper sequence), respectively.}
\end{figure}
Fig.~\ref{fig:CredibileLevel} takes the full posterior distribution of the events assigned to the groups into account. There are 500 fits done in equal step sizes of the credible intervals of the distributions in total mass and chirp mass. Please note, the data shown in Fig.~\ref{fig:CredibileLevel} should be compared to the values given in Table~\ref{tab:fits_weighting}.
%The reduced $\chi^2$ is normalized to the values shown in Table~\ref{tab:fits}.
For group~1 there are strong differences here, because the full posterior include the second peaks at low redshift\footnote{\citet{GWTC3} did not state the detailed cut criteria which they applied to remove the second peak in the distributions to obtain the values in their Table IV. Hence, the two events GW191219\_163120 and GW200322\_091133 have by definition a published posterior distribution which is inconsistent with the quoted final values in the table.}.
%Nevertheless, it is clearly visible that all data points with an acceptable reduced $\chi^2$ are still far of the expectation from channels of isolated binary evolution. Additionally, 
When including the wings of the posteriors the uncertainties on the parameters get larger. It is evident that only the distribution close to the most likely values (low credible interval values) support linear relations well. While the full posteriors do not give a strong support to the relations fitted here, parts of them (the most probable ones) do. Hence, those relations may only provide a failed check on the Bayesian interface used to obtain the posterior distributions. It should be noted, that passing several of those kind of checks does not prove the Bayesian method, it can only be disproven by failing one check \citep{rtl22}.

\begin{figure}
 \centering
 \includegraphics[width=\columnwidth, clip, trim=10px 10px 20px 20px]{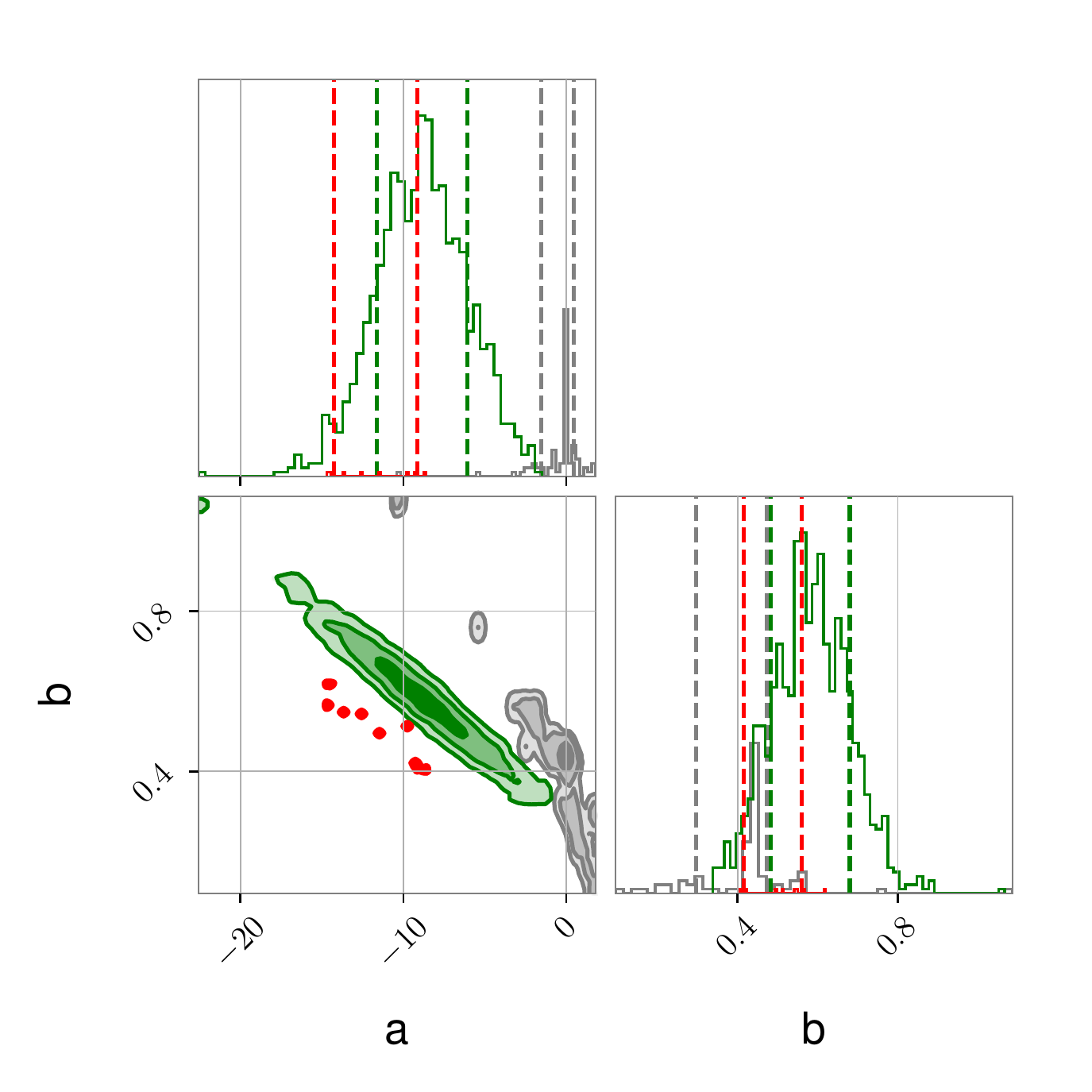}
 \caption{\label{fig:corner}The fit parameter distribution obtained via bootstrapping for the full event posteriors, where fits with reduced $\chi^2>1$ are rejected. Groups~1 and 2 are shown in red and green, respectively. For comparison the BPS merger channel fits are shown in gray. It should be noted that the gray distribution does not take the counts of the different channels into account.}
\end{figure}
Another approach to use the full event posteriors is to determine the fitting parameters via random sampling of the posteriors. The results of 10\,000 trials are shown Fig.~\ref{fig:corner}. The distributions in Fig.~\ref{fig:corner} show only data, where the reduced $\chi^2\leq 1$, hence the fit is expectable as a sufficiently good fit. There are still no overlaps between the different distributions. Only if one includes results of fits which indicate to be a worse linear fit one can obtain overlaps in the parameter distributions. The small number of expectable fits for group~1 are caused by the fact that group~1 contains events with multi-peak posteriors and/or posteriors which cover a large part of the parameter space.
%It should be noted that Fig.~\ref{fig:corner} do not take the goodness of the fits into account. Most parts in the wings of the distributions provide results which should have been rejected as a good fit. Therefore, one should not understand the overlapping regions of the different distributions as consistencies. Because those overlaps are vastly dominated by results, which should be rejected, but are included here for completeness.
All in all, it again points out that only a part of the posterior distributions support the relations here. If the posteriors are free of biases and/or misspecification no part should support such relations, especially not the most likely parts, cf. Fig.~\ref{fig:CredibileLevel}.

\begin{figure}
 \centering
 \includegraphics[width=\columnwidth]{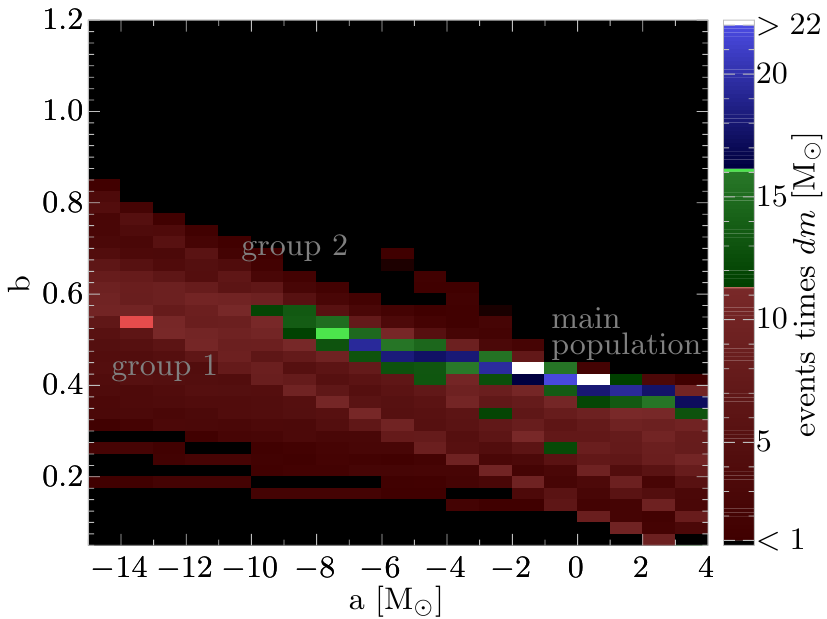}
 \includegraphics[width=\columnwidth]{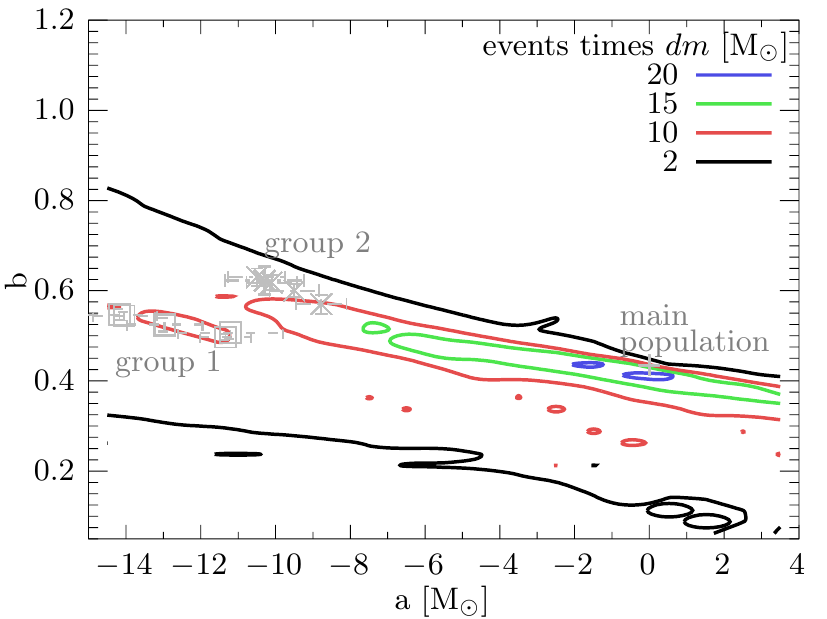}
 \caption{\label{fig:linear3}The top panel is similar to Fig.~\ref{fig:linear2} but using all data points within the 90\% credible interval for each event instead of the median values only. The bottom panel shows the contours to better view the local maxima, which represent the groups.}
\end{figure}
With taking all the data within the 90\% credible interval of the posteriors into account Fig.~\ref{fig:linear2} changes to Fig.~\ref{fig:linear3}. As expected form earlier considerations of the full posteriors this blurs out the parts which do not belong to the main population with a mass ratio close to 1. Nevertheless, there is still a local maxima visible for group~1. Group~2 becomes more and more contaminated by the main population because of several events being in the overlapping region between the relation of group~2 and the population with a mass ratio close to 1.

\begin{table}
 \centering
 \caption{\label{tab:members1}Overlaps of the full posteriors with the relation derived for group~1. Above the horizontal line all events with an overlap $>50\%$ to the $2\sigma$ region of the relation are shown, below are the systems we assigned to the group additionally. The columns from left to right are: 1) event name, 2) our group assignment, 3) percentage of overlap with the $1\sigma$ region of the relation, 4) percentage of overlap with the $2\sigma$ region of the relation [sorted by]. The inequalities in columns 3 and 4 are used to indicate the expected bias of the stated values.}
 %, 5) indicating the bias from over- to underestimated [$\uparrow$, $\nearrow$, $\rightarrow$, $\searrow$, $\downarrow$].
 \begin{tabular}{cccc}
  \hline
  event & g & $p(e|g_{1,1\sigma})$ & $p(e|g_{1,2\sigma})$\\
  \hline
  GW191219\_163120 & 1\phantom{0} & $\approx 82.6\%$ & $\approx 98.0\%$\\%
  GW191109\_010717 & 1b & $\lesssim 49.8\%$ & $\lesssim 93.8\%$\\
  GW190602\_175927 & 1b & $\lesssim 59.5\%$ & $\lesssim 93.4\%$\\
  GW190519\_153544 & 1c & $< 53.2\%$ & $< 92.0\%$\\
  GW190706\_222641 & 1c & $\lesssim 48.6\%$ & $\lesssim 82.6\%$\\
  GW190521\phantom{\_000000} & 1a & $\lesssim 42.1\%$ & $\lesssim 82.5\%$\\
  GW200220\_061928 & 1a & $\lesssim 51.5\%$ & $\lesssim 80.3\%$\\
  GW190620\_030421 & - & $< 23.4\%$ & $< 55.3\%$\\
  \hline
  GW190929\_012149 & 1\phantom{0} & $\gtrsim 14.4\%$ & $\gtrsim 30.2\%$\\
  GW191127\_050227 & 1\phantom{0} & $> 13.2\%$ & $> 26.4\%$\\
  GW200322\_091133 & 1\phantom{0} & $\gtrsim \phantom{0}5.8\%$ & $\gtrsim 14.5\%$\\
  GW200208\_222617 & 1\phantom{0} & $> \phantom{0}4.6\%$ & $> \phantom{0}9.3\%$\\
  \hline
 \end{tabular}
\end{table}

\begin{table}
 \centering
 \caption{\label{tab:members2}Overlaps of the full posteriors with the relation derived for group~2. Above the horizontal line all events with an overlap $>50\%$ to the $1\sigma$ region of the relation are shown, below are the systems we assigned to the group additionally. The columns from left to right are: 1) event name, 2) our group assignment, 3) percentage of overlap with the $1\sigma$ region of the relation [sorted by], 4) percentage of overlap with the $2\sigma$ region of the relation. The inequalities in columns 3 and 4 are used to indicate the expected bias of the stated values.}
 \begin{tabular}{cccc}
  \hline
  event & g & $p(e|g_{2,1\sigma})$ & $p(e|g_{2,2\sigma})$\\
  \hline
  GW190814\phantom{\_000000} & 2- & $\approx 99.5\%$ & $\approx 100.0\%$\\
  GW170814\phantom{\_000000} & 2++ & $< 97.8\%$ & $< 100.0\%$\\
  GW190828\_063405 & 2+++ & $< 92.8\%$ & $< 100.0\%$\\
  GW170104\phantom{\_000000} & 2+ & $< 92.4\%$ & $< \phantom{0}99.3\%$\\
  GW200311\_115853 & - & $< 80.5\%$ & $< \phantom{0}99.5\%$\\
  GW190630\_185205 & - & $< 76.2\%$ & $< \phantom{0}98.7\%$\\
  GW170809\phantom{\_000000} & - & $< 73.7\%$ & $< \phantom{0}97.0\%$\\
  GW190412\phantom{\_000000} & 2- & $< 70.3\%$ & $< \phantom{0}97.7\%$\\
  GW170818\phantom{\_000000} & - & $< 70.1\%$ & $< \phantom{0}98.3\%$\\
  GW200129\_065458 & - & $< 69.5\%$ & $< \phantom{0}95.9\%$\\
  GW190915\_235702 & - & $< 66.9\%$ & $< \phantom{0}95.0\%$\\
  GW191204\_110529 & 2+ & $< 66.0\%$ & $< \phantom{0}95.7\%$\\
  GW190413\_052954 & 2+++ & $< 64.7\%$ & $< \phantom{0}92.3\%$\\
  GW200112\_155838 & - & $< 62.1\%$ & $< \phantom{0}99.2\%$\\
  GW190513\_205428 & - & $< 59.5\%$ & $< \phantom{0}88.9\%$\\
  GW200210\_092254 & 2- & $\approx 57.7\%$ & $\approx \phantom{0}83.3\%$\\
  GW200209\_085452 & - & $< 57.5\%$ & $< \phantom{0}94.6\%$\\
  GW190527\_092055 & - & $< 53.1\%$ & $< \phantom{0}81.3\%$\\
  GW190719\_215514 & - & $< 51.7\%$ & $< \phantom{0}77.4\%$\\
  GW150914\phantom{\_000000} & - & $< 50.9\%$ & $< 100.0\%$\\
%  GW190803\_022701 & - & $< 45.2\%$ & $< \phantom{0}91.3\%$\\
%  GW191215\_223052 & - & $< 44.2\%$ & $< \phantom{0}98.5\%$\\
%  GW200219\_094415 & - & $< 42.4\%$ & $< \phantom{0}92.0\%$\\
%  GW200208\_130117 & - & $< 39.1\%$ & $< \phantom{0}92.3\%$\\
%  GW190408\_181802 & - & $< 37.9\%$ & $< \phantom{0}99.9\%$\\
%  GW190727\_060333 & - & $< 32.7\%$ & $< \phantom{0}91.5\%$\\
%  GW200224\_222234 & - & $< \phantom{0}2.2\%$ & $< \phantom{0}94.1\%$\\
  \hline
  GW200306\_093714 & 2 & $< 48.6\%$ & $< \phantom{0}84.0\%$\\
  GW191113\_071753 & 2- & $\approx 30.4\%$ & $\approx \phantom{0}59.5\%$\\
  \hline
 \end{tabular}
\end{table}

In Tables~\ref{tab:members1} and \ref{tab:members2} we show the fraction of the posterior overlapping with the relations derived for group~1 and 2, respectively. We need to caution the reader that those values cannot be used as agreements to the groups to invoke a group assignment based on them. There are several systematic effects at work when deriving those values. First, a broad posterior cannot have a high percentage in agreeing with a small area of a tight relation. Thus any event with a larger uncertainty on its values than the uncertainty of the relation may cause a too small agreement with the relation. On the other hand, events with very small uncertainties may get a too good agreement when they sole fall into the outer parts of the relation uncertainty. We use the square route of the 90\% credible area in units of the mean of all events to estimate whether an event is effected by this bias. Second, because of the broadening of the relations toward higher masses, events with higher masses in general are favoured against low mass events. We use the mean binary mass of the group over the mean binary mass of each event to estimate whether an event is effected by this bias. Third, all data points close to the limiting mass ratio of 1, will have clipping effects imprinted in the posteriors leading to higher overlaps if they are close to the relations. We use the mean $dm$ of an event in units of the mean $dm$ of the group to estimate whether an event is effected by this bias. It is unclear whether the ratios indicating the biases fully compensate or may even overcompensate some. Especially, the third bias measure has the tendency to introduce a new bias. Thus, a quantitative usage of those is probably biased, too. Therefore, we derive only a qualitative statement about the bias from them instead of providing bias corrected values because they would not be bias free nor could be ensured that they would be always less biased. To indicate the bias we used inequality signs, depending how many of the three criteria apply to an event and the combined bias (using a simple product here). It should be noted, that mainly the systems which are assigned clearly by eye inspection get too often underestimated caused by biases when using an automatic assignment based on the overlap of the posteriors and relations.

At this point it is already unexpected that any further analysis taking the full posterior distributions into account will strongly support the relations. This is caused by the broad distributions of the individual events. While the real event would have a single value of each characteristics the posteriors are broader because of the point spread functions of the instruments which are not fully compensated by the Bayesian analysis to determine the posteriors nor can be excluded that some parts of the posterior are newly introduced during this analysis.
%Therefore, we omit another Bayesian analysis which might introduce additional biases without a robust measure of them as long as the true values are unknown.
We still omit to use the exactly same methodology as used to derive the posteriors to avoid the fact, that a disapproval of the relations cannot be differentiated from a passed self-consistency check if the nature of the relations is introduced by the methodology and their assumptions used for the data analysis.

%\section{Bayesian ...}
%\textbf{Tbc.}

\section{Stochastic Check}\label{sec:stochastic}
\begin{figure}
 \centering
 \includegraphics[width=\columnwidth]{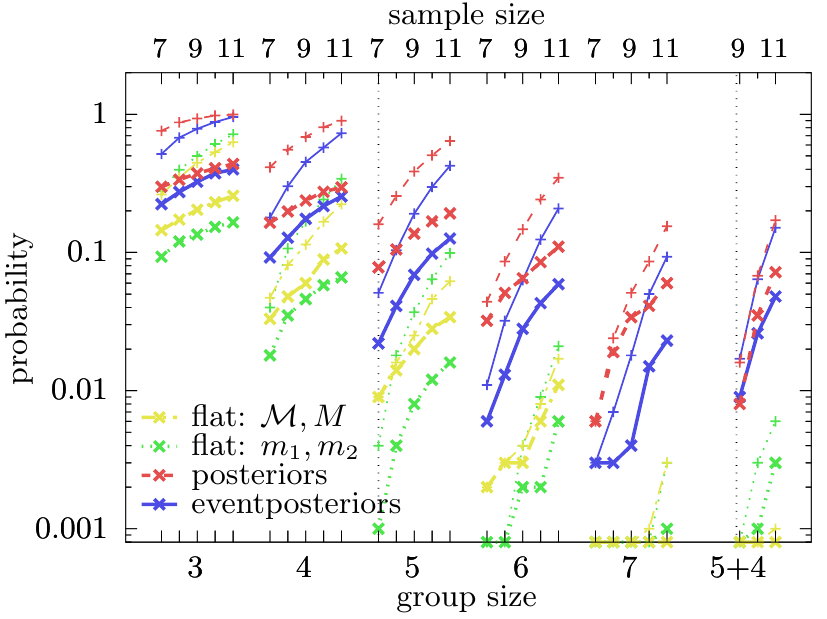}
 \caption{\label{fig:Random}The probability to find a linear relation with a reduced $\chi^2\leq1$ for a group of a given size out of a sample with a given number of events with largest $dm$ out of 86 events in total. The color/line style indicate the underlying population to be flat in the chirp and total mass (yellow, dash dotted), flat in the component masses (green, dotted), flat in all posterior points (red, dashed), and flat in the events and their individual posterior (blue, solid). The pluses (thin lines) mark an absolutely random combination, while the crosses (thick lines) require, the event with the largest $dm$ to be part of the group. The requirements to claim groups in this paper are marked with the vertical black dotted lines.}
\end{figure}
In this section we investigate the possibility of a random nature of the relations. The fact, that there is no unique way to determine the subpopulations automatically, requires a more generous approach here. We use different underlying populations to get 86 events from them and sort them according to the value of $dm$. We use four different assumed underlying distributions: First, we sample uniformly from the observed 86 events and within those we sample from the posterior data points. Second, we sample uniformly from the full sample of posterior points. This gives emphasis on events with more densely sampled posteriors. Third, we sample flat in the component masses $m_1$ and $m_2$ up to $\unit{100}{\Msun}$ for $m_1$ and up to $m_1$ for $m_2$. This is similar to the prior used in the analysis of the observational data. Fourth, we sample flat in the parameter space we are looking at (the chirp and total mass). For each of the $1\,000$ trials per distribution we take the events with the largest $dm$ to fit linear relations to them. The number of considered events out of the 86 drawn ones with largest $dm$ is indicated as sample size (top x-axis between 7 and 11 events) in Fig.~\ref{fig:Random}. As in the observational data set (see Table~\ref{tab:data}) we allow for skipping some data points. Hence, we specify a group size in the range 3 to 7. For this number of group members we fit all possible combinations and record how many satisfy a reduced $\chi^2\leq1$ for an uncertainty of $\unit{1}{\Msun}$ in the chirp mass, like in Table~\ref{tab:fits}. Fig.~\ref{fig:Random} shows how often in the $1\,000$ trials at least one acceptable fit is achieved (the values below $0.001$ represent combinations of group and sample size with zero fits passing the threshold). Finally, there are fits, with two simultaneous groups of size 5 and 4 with no common events between the groups like for groups (1 and 2-).

Group sizes of 3 and 4 do show significant contributions, where a linear fit could just be fortuitous. That is a confirmation of the reasoning to require at least 5 data points per relation in this work. The two combinations of group and sample size like we have by the observational sample are marked with the black dotted vertical lines. The crosses (thick lines) do indicate the probability with the additional requirement, that the event with the largest $dm$ has to be part of the fitted group (in the larger one in the case of two groups). Thus, the crosses are always below or equal the pluses. When taking the two events reported by \citet{OGC3,OGC4} with a high $dm$ into account the sample size of the marked combination would be larger by 2. Nevertheless, the crosses are still below $10\%$, showing that a pure random nature of the relations is outside a $90\%$ confidence. Still it cannot be excluded that the relations get less support with future observations. It should be noted, that we choose the thresholds here well above the values obtained for the here presented relations. Additionally, we do not require the relations to be in a certain area of the $a$ vs. $b$ plane. Therefore, the values in Fig.~\ref{fig:Random} should be understood as upper limits. Interestingly, the random samples based on the posteriors do show a significantly larger potential of getting linear relations by chance. This indicates, that the posteriors already give more support to linear relations than the priors.

\end{document}